\documentstyle[12pt]{article}

\newcommand \beq{\begin{eqnarray}}
\newcommand \eeq{\end{eqnarray}}
\begin{document}

\begin{titlepage}
\begin{flushright}

{Saclay-T94/013}
\end{flushright}
\vspace*{3cm}
\begin{center}
\baselineskip=13pt
{\Large NON-ABELIAN PLANE-WAVES IN THE\\QUARK-GLUON PLASMA\\}
\vskip0.5cm
Jean-Paul BLAIZOT\footnote{CNRS}  and
Edmond IANCU\\
{\it Service de Physique Th\'eorique\footnote{Laboratoire de la Direction des
Sciences de la Mati\`ere du Commissariat \`a l'Energie
Atomique}, CE-Saclay \\ 91191 Gif-sur-Yvette, France}\\
\vskip0.5cm
January 1994
\end{center}
\baselineskip=22pt
\vskip 2cm
\begin{abstract}
We present new, non-abelian, solutions to the equations of
motion which describe the collective excitations of a quark-gluon plasma
at high temperature. These solutions correspond to longitudinal and transverse
plane-waves propagating through the plasma.
\end{abstract}
\vskip 4cm

\begin{flushleft}
Submitted to Physics Letters {\bf B}
\end{flushleft}

\end{titlepage}

\def\bfepsilon{\mbox{\boldmath$\epsilon$}}
\def\bfcalA{\mbox{\boldmath${\cal A}$}}
\def\bfcalS{\mbox{\boldmath${\cal S}$}}

\def\square{\hbox{{$\sqcup$}\llap{$\sqcap$}}}   
\def\grad{\nabla}                               
\def\del{\partial}                              

\def\frac#1#2{{#1 \over #2}}
\def\smallfrac#1#2{{\scriptstyle {#1 \over #2}}}
\def\half{\ifinner {\scriptstyle {1 \over 2}}
   \else {1 \over 2} \fi}

\def\bra#1{\langle#1\vert}              
\def\ket#1{\vert#1\rangle}              

\def\simge{\mathrel{%
   \rlap{\raise 0.511ex \hbox{$>$}}{\lower 0.511ex \hbox{$\sim$}}}}
\def\simle{\mathrel{
   \rlap{\raise 0.511ex \hbox{$<$}}{\lower 0.511ex \hbox{$\sim$}}}}


\def\parenbar#1{{\null\!                        
   \mathop#1\limits^{\hbox{\fiverm (--)}}       
   \!\null}}                                    
\def\nunubar{\parenbar{\nu}}
\def\ppbar{\parenbar{p}}


\def\buildchar#1#2#3{{\null\!                   
   \mathop#1\limits^{#2}_{#3}                   
   \!\null}}                                    
\def\overcirc#1{\buildchar{#1}{\circ}{}}


\def\slashchar#1{\setbox0=\hbox{$#1$}           
   \dimen0=\wd0                                 
   \setbox1=\hbox{/} \dimen1=\wd1               
   \ifdim\dimen0>\dimen1                        
      \rlap{\hbox to \dimen0{\hfil/\hfil}}      
      #1                                        
   \else                                        
      \rlap{\hbox to \dimen1{\hfil$#1$\hfil}}   
      /                                         
   \fi}                                         %


\def\subrightarrow#1{
  \setbox0=\hbox{
    $\displaystyle\mathop{}
    \limits_{#1}$}
  \dimen0=\wd0
  \advance \dimen0 by .5em
  \mathrel{
    \mathop{\hbox to \dimen0{\rightarrowfill}}
       \limits_{#1}}}                           

\def\real{\mathop{\rm Re}\nolimits}     
\def\imag{\mathop{\rm Im}\nolimits}     

\def\tr{\mathop{\rm tr}\nolimits}       
\def\Tr{\mathop{\rm Tr}\nolimits}       
\def\Det{\mathop{\rm Det}\nolimits}     

\def\mod{\mathop{\rm mod}\nolimits}     
\def\wrt{\mathop{\rm wrt}\nolimits}     


\def\TeV{{\rm TeV}}                     
\def\GeV{{\rm GeV}}                     
\def\MeV{{\rm MeV}}                     
\def\KeV{{\rm KeV}}                     
\def\eV{{\rm eV}}                       

\def\mb{{\rm mb}}                       
\def\mub{\hbox{$\mu$b}}                 
\def\nb{{\rm nb}}                       
\def\pb{{\rm pb}}                       

%
\def\journal#1#2#3#4{\ {#1}{\bf #2} ({#3})\  {#4}}

\def\AdvPhys{\journal{Adv.\ Phys.}}
\def\AnnPhys{\journal{Ann.\ Phys.}}
\def\EurophysLett{\journal{Europhys.\ Lett.}}
\def\JApplPhys{\journal{J.\ Appl.\ Phys.}}
\def\JMathPhys{\journal{J.\ Math.\ Phys.}}
\def\LettNuovoCimento{\journal{Lett.\ Nuovo Cimento}}
\def\Nature{\journal{Nature}}
\def\NPA{\journal{Nucl.\ Phys.\ {\bf A}}}
\def\NPB{\journal{\it {Nucl.\ Phys.\ {\bf B}}}}
\def\NuovoCimento{\journal{Nuovo Cimento}}
\def\Physica{\journal{Physica}}
\def\PLA{\journal{Phys.\ Lett.\ {\bf A}}}
\def\PLB{\journal{\it {Phys.\ Lett.\ {\bf B}}}}
\def\PhysRev{\journal{Phys.\ Rev.}}
\def\PRC{\journal{Phys.\ Rev.\ {\bf C}}}
\def\PRD{\journal{\it {Phys.\ Rev.\ {\bf D}}}}
\def\PRL{\journal{\it {Phys.\ Rev.\ Lett.}}}
\def\PhysRept{\journal{Phys.\ Repts.}}
\def\ProcNatlAcadSci{\journal{Proc.\ Natl.\ Acad.\ Sci.}}
\def\ProcRoySoc{\journal{Proc.\ Roy.\ Soc.\ London Ser.\ A}}
\def\RevModPhys{\journal{\it {Rev.\ Mod.\ Phys.}}}
\def\Science{\journal{Science}}
\def\SovPhysJETP{\journal{\it {Sov.\ Phys.\ JETP}}}
\def\SovPhysJETPLett{\journal{Sov.\ Phys.\ JETP Lett.}}
\def\SovJNuclPhys{\journal{\it {Sov.\ J.\ Nucl.\ Phys.}}}
\def\SovPhysDoklady{\journal{Sov.\ Phys.\ Doklady}}
\def\ZPhys{\journal{Z.\ Phys.}}
\def\ZPhysA{\journal{Z.\ Phys.\ A}}
\def\ZPhysB{\journal{Z.\ Phys.\ B}}
\def\ZPhysC{\journal{Z.\ Phys.\ C}}

\baselineskip=22pt

\parindent=20pt

\setcounter{equation}{0}
\section{Introduction. General equations}

The long wavelength excitations of a quark-gluon plasma  are
 collective excitations which are described by nonlinear equations
generalizing the classical Yang-Mills equations in the vacuum.
Most studies have been limited so far to the weak field limit, where the
equations become linear and the excitations reduce  to abelian-like
plasma waves\cite{Klimov81}, very much similar to the electromagnetic
waves in ordinary plasmas\cite{Silin60}. The
 purpose of this paper is to present new, truly
non-abelian, plane-wave solutions that we have obtained recently.
At leading order in the gauge coupling
$g$, (we assume $g\ll 1$ in the high temperature, deconfined, QCD plasma),
the collective dynamics is entirely described by a set of  equations
for the gauge mean fields $A^\mu_a(x)$ which describe the long wavelength
($\lambda\sim 1/gT$) and low frequency ($\omega\sim gT$) excitations
 ($T$ denotes the temperature)\cite{us,Pisarski}.
(Throughout this work, the  greek indices refer to Minkovski space, while the
latin ones are color indices for the adjoint representation of the
gauge group $SU(N)$.)  The equations satisfied by  $A^\mu_a(x)$ are
\beq
\label{ava}
\left [\, D^\nu,\, F_{\nu\mu}(x)\,\right ]_a
\,=\,j^{\mu}_a(x),
\eeq
where $D^\mu = \del^\mu+igA^\mu(x)$, $A^\mu= A^\mu_a T^a$,
and  $F^{\mu\nu}= [D^\mu, D^\nu]/(ig) = F^{\mu\nu}_aT^a$.
In the right hand side,  $j^\mu_a$   is the {\it induced current}
which describes the  response of the plasma to the
color gauge fields $A_a^\mu$. (We do not consider here any external color
source.) Its expression is\cite{us,EMT}
\beq\label{j1}
j^{\mu}_a(x)\,=\,3\,\omega^2_p\int\frac{d\Omega}{4\pi}
\,v^\mu\,W_a^0(x;v),\eeq
where $\omega^2_p\equiv (2N+N_{\rm f})g^2 T^2/18$  is the {\it plasma
 frequency},  $v^\mu\equiv (1,\,{\bf v})$, ${\bf v}\equiv
{\bf k}/k$, $k\equiv |{\bf k}|$, and the  integral $\int
d\Omega$ runs over all the directions of the unit vector ${\bf v}$.
The functions  $W^\mu\equiv W_a^\mu T^a$ are defined as the solutions to
\beq\label{eqw}
\left[ v\cdot D_x,\, W^\mu(x;v)\right]\,=\,F^{\mu\rho}(x)\,v_\rho.\eeq
For retarded conditions  (i.e., $A_a^\mu(x)\to 0$ as  $x_0\to -\infty$)
\beq\label{W}
W^\mu_a(x;v)\,=\, \int_0^\infty du\, U_{ab}(x,x-vu)\, F_b^{\mu\rho}(x-vu)
\,v_\rho,\eeq
where  $U(x,y)$ is the parallel transporter along the straight line
$\gamma$   joining $x$ and $y$:
\beq\label{pt} U(x,y)=P\exp\{ -ig\int_\gamma dz^\mu A_\mu(z)\}.\eeq
The energy-momentum tensor of an arbitrary gauge field configuration in
the plasma has been recently computed\cite{EMT}, with the following results
 for the energy density
\beq\label{enden}
T^{00}(x)\,=\,\frac{1}{2}\Bigl({\bf E}^a(x)\cdot{\bf E}^a(x)\,+\,
{\bf B}^a(x)\cdot{\bf B}^a(x)\Bigr)\,+\,\frac{3}{2}\,
\omega^2_p\int\frac{d\Omega}{4\pi}\,W_a^0(x;v)\,W_a^0(x;v),\eeq
and for the energy flux density, or Poynting vector,
$S^i\equiv T^{i0}$,
\beq\label{Poyn}
{\bf S}(x)\,=\,{\bf E}^a(x)\times{\bf B}^a(x)\,+\,
\frac{3}{2}\,
\omega^2_p\int\frac{d\Omega}{4\pi}\,\,{\bf v}\,\,W_a^0(x;v)\,W_a^0(x;v).\eeq
The non-abelian field strengths are, as usually, $E^i\equiv F^{i0}$ and
$B^i_a\equiv -(1/2)\epsilon^{ijk}\,F^{jk}_a$. The first terms in the r.h.s.
of eqs.~(\ref{enden})--(\ref{Poyn})
 represent the standard Yang-Mills contributions;
 the terms depending on $W^0_a(x;v)$ are related to the color
polarizability of the plasma.
In the absence of external sources, the energy conservation requires that
\beq\label{EP}
\del_0\,T^{00}(x)\,+\,\del_i\,S^i(x)\,=\,0.\eeq
It can be easily verified  by using the equations of
motion (\ref{ava})--(\ref{eqw}) that this equation is indeed satisfied.

\setcounter{equation}{0}
\section{Plane-wave solutions}

 In this letter, we study particular  plane-wave solutions
which  depend on $x^\mu$ only through the variable $z\equiv p^\mu x_\mu$, where
 $p^\mu\equiv (\omega, {\bf p})$ is a fixed, time-like, four-vector
($p^\mu p_\mu=\omega^2-p^2>0$, $p\equiv |{\bf p}|$).
Thus, we search for solutions of the form
\beq\label{pw}
A^\mu_a(x)\,=\,{\cal A}^\mu_a(p\cdot x),\eeq
in a gauge to be specified later. For the classical Yang-Mills
equations in the vacuum, plane-wave solutions of this  form have been
investigated in Refs. \cite{Ishida75,Baseyan79}. Another class
of non-abelian plane-waves in the vacuum was considered by
 Coleman\cite{Coleman77}. We note that, in contrast to the vacuum case,
the solutions of eqs.~(\ref{ava}) for the high temperature plasma
have direct physical relevance: they correspond
 to the collective color excitations of the QCD plasma.

In deriving eqs.~(\ref{ava})--(\ref{eqw}), we have assumed\cite{us} the
 gauge fields  to be weak ($A\simle T$) and slowly varying ($\del A\sim gTA$).
For consistency, we therefore require that $\omega\sim p\sim gT$ and
$|{\cal A}^\mu_a|\simle T$ for the plane waves (\ref{pw}).
An important consequence of the Ansatz (\ref{pw}) is that the corresponding
functions   $W_a^\mu(x;v)$  are local and linear in the gauge fields:
\beq\label{pW}
W_a^\mu(x;v)\,=\,-\,{\cal A}_a^\mu(z)\,+\,\frac{p^\mu}{v\cdot p}\,
v\cdot{\cal A}_a(z),\eeq
as can be verified on eq.~(\ref{eqw}). Then, the induced current (\ref{j1}) may
 be easily computed, with the following results
\beq\label{dj}
\rho_a(x)\,=\,-\,3\omega_p^2 \,\alpha (\omega/p)\Bigl(
 {\cal A}^L_a\,-\,\frac{p}{\omega}\,{\cal A}_a^0\Bigr),\eeq
and \beq\label{pj}
{\bf j}_a(x)\,=\,-\,3\omega_p^2 \,\frac{\omega}{p}\biggl\{
\alpha (\omega/p)\Bigl({\cal A}^L_a\,-\,\frac{p}{\omega}\,{\cal A}_a^0\Bigr)
\,{\hat {\bf p}}\,-\,\beta(
\omega/p)\,{\bfcalA}_a^T\biggr\},\eeq
where   ${\cal A}_a^L\equiv{\hat {\bf p}}\cdot {\bfcalA}_a$
(${\hat {\bf p}}\equiv {\bf p}/p$), and ${\bfcalA}_a^T\equiv
{\bfcalA}_a-{\hat {\bf p}}{\cal A}^L_a$. The functions $\alpha (u)$ and
 $\beta (u)$ ($u\equiv  \omega/p$) appear when evaluating the angular
 integral in (\ref{j1}). We have ($\omega > p$)
\beq\label{vivj}
\int\frac {d\Omega}{4\pi}\,\,\frac {v^i\,v^j}
{\omega - {\bf v}\cdot{{\bf p}}}\,=\,\frac{ \alpha ({\omega}/{p})}{p^2}\,
\hat p^i\,\hat p^j\,+\,\frac{\beta ({\omega}/{p})}{p^2}\,
(\delta^{ij}-\hat p^i\,\hat p^j),\eeq
with
\beq\label{albe}
\alpha(u)\equiv u\,\Bigl(Q(u)\,-\,1\Bigr),\qquad\,\,
\beta(u)\equiv \frac{u}{2}\,\Bigl(1\,-\,\frac{u^2-1}{u^2}\,Q(u)\Bigr),\eeq
and $Q(u)$ is defined by
\beq\label{Q}
Q(u)\equiv \frac{u}{2}\,\ln\frac{u+1}{u-1}.\eeq
For $u>1$, the functions $\alpha(u)$ and $\beta(u)$ are positive.
Remark that the current (\ref{dj})--(\ref{pj}) is  not only covariantly
 conserved, $[D_\mu, j^\mu]=0$,  as it should for the consistency of
 eq.~(\ref{ava}), but it also verifies $\del_\mu j^\mu_a=p_\mu j^\mu_a=0$.

The expressions (\ref{dj})--(\ref{pj}) for the induced current are formally
identical to those corresponding to an abelian plasma, that is, they
involve only the gauge field polarization tensor, and no higher order
 irreducible amplitudes. This simplification represents
 an essential feature of the plane-wave fields that we are considering here.

For later use, we  evaluate here the polarization pieces of the energy
 density (\ref{enden})  and of the Poynting vector (\ref{Poyn})
for the plane-waves (\ref{pw}). We need the  integrals ($\omega > p$):
\beq\label{vivj2}
\int\frac {d\Omega}{4\pi}\,\,\frac {v^i\,v^j}
{(\omega - {\bf v}\cdot{{\bf p}})^2}\,=\,\frac{ a({\omega}/{p})}{p^2}\,
\hat p^i\,\hat p^j\,+\,\frac{b({\omega}/{p})}{p^2}\,
(\delta^{ij}-\hat p^i\,\hat p^j),\eeq
and
\beq\label{vivjvk2}
\int\frac {d\Omega}{4\pi}\,\,\frac {v^i\,v^j\,v^k}
{(\omega - {\bf v}\cdot{{\bf p}})^2}\,=\,\frac{ c({\omega}/{p})}{p^2}\,
\hat p^i\,\hat p^j\,\hat p^k\,+\,\frac{d({\omega}/{p})}{p^2}\,
(\delta^{jk}\,\hat p^i\,+\,\delta^{ik}\,\hat p^j\,+\,\delta^{ij}\,\hat p^k),
\eeq  where
\beq\label{ab}
{a}(u)\equiv 1\,+\,\frac{u^2}{u^2-1}\,-\,2Q(u),\qquad\,\,
{b}(u)\equiv Q(u)\,-\,1,\eeq
and
\beq\label{cd}
d(u)\,\equiv \,\frac{3}{2}\,u\,\Bigl( Q(u)\,-\,1\Bigr)\,-\,\frac{1}{2u}\,Q(u),
\nonumber\\
c(u)\,+\,3\,d(u)\,\equiv \, \frac{u}{u^2-1}\Bigl[1\,-\,3\,(u^2-1)\,
\Bigl( Q(u)\,-\,1\Bigr)\Bigr],\eeq
with $Q(u)$  given by (\ref{Q}).
 For $u>1$, the functions $a(u)$ and  $b(u)$ are both
 positive. By using these results, together with eq.~(\ref{pW}), we get
\beq\label{ennon}
T^{00}(x)\,=\,\frac{1}{2}\Bigl({\bf E}^a\cdot{\bf E}^a\,+\,
{\bf B}^a\cdot{\bf B}^a\Bigr)\,+\, \frac{3}{2}\,\omega^2_p\,
\,\frac{\omega^2}{p^2}\biggl\{
a (\omega/p)\Bigl({\cal A}^L_a\,-\,\frac{p}{\omega}\,{\cal A}_a^0\Bigr)^2
\,+\,b(\omega/p)\,{\bfcalA}_a^T\cdot {\bfcalA}_a^T\biggr\},\nonumber\\ \eeq
and
\beq\label{Pnon}
{\bf S}(x)\,=\,{\bf E}^a\times{\bf B}^a &+& \frac{3}{2}\,\omega^2_p\,
\,\frac{\omega^2}{p^2}\biggl\{ \Bigl(c(\omega/p)\,+\,3\,
d (\omega/p)\Bigr)\Bigl({\cal A}^L_a\,-\,\frac{p}{\omega}\,{\cal A}_a^0\Bigr)^2
\,\hat {\bf p}\nonumber\\
&+&d(\omega/p)\,{\bfcalA}_a^T\cdot {\bfcalA}_a^T\,\hat {\bf p}\,+\,
2\,d(\omega/p)\Bigl({\cal A}^L_a\,-\,\frac{p}{\omega}\,{\cal A}_a^0\Bigr)\,
{\bfcalA}_a^T\biggr\}.\eeq

In the rest of this letter, we work in the covariant gauge $p_\mu A^\mu_a=0$.
With the Ansatz (\ref{pw}), the gauge field equations (\ref{ava}) become then
\beq\label{pweq}
p_\nu p^\nu\,{\ddot {\cal A}}^\mu\,+\,i\,g\,p^\mu\,\Bigl [{\cal A}_\nu, \dot
{\cal A}^\nu\Bigr ]\,-\,g^2\,\biggl [{\cal A}_\nu, \Bigl [{\cal A}^\nu,
{\cal A}^\mu \Bigr ]\biggr ]\,=\,j^\mu.\eeq
(Throughout, the overdots indicate derivatives with
respect to the argument of the function, here  $z$.) When contracted
 with $p_\mu$, eq.~(\ref{pweq}) reduces to
\beq\label{const}
\Bigl [{\cal A}_\nu, \dot {\cal A}^\nu\Bigr ]\,=\,0,\eeq
(recall that $p^\nu p_\nu>0$ and $p^\mu j_\mu^a=0$). Therefore, we need only
consider the simplified equation
\beq\label{pweq1}
p_\nu p^\nu\,{\ddot {\cal A}}^\mu\,-\,g^2\,\biggl [{\cal A}_\nu, \Bigl [{\cal
A}^\nu,
{\cal A}^\mu \Bigr ]\biggr ]\,=\,j^\mu,\eeq
together with the constraint (\ref{const}).
In the present gauge, $\omega {\cal A}_a^0=p  {\cal A}_a^L$, and the induced
current (\ref{dj})--(\ref{pj}) can be rewritten as
\beq\label{ind}
\rho_a&=&-\,\Omega_L^2\, {\cal A}_a^0,\nonumber\\
{\bf j}_a&=&-\,\Omega_L^2\, {\cal A}_a^L\,\hat{\bf p}\,-\,
\Omega_T^2\, \bfcalA_a^T,\eeq
where
\beq\label{OL}
\Omega_L^2\,\equiv\,3\,\omega_p^2\,\frac{\omega^2-p^2}
{p^2}\,\biggl(Q\biggl(\frac{\omega}{p}\biggr)
\,-\,1\biggr),\eeq
and
\beq\label{OT}
\Omega_T^2\,\equiv\,\frac{3}{2}\,\omega_p^2
\frac{\omega^2}{p^2}\,\biggl(1
\,-\,\frac{\omega^2-p^2}{\omega^2}Q\biggl(\frac{\omega}{p}\biggr)\biggr).\eeq
Since $\omega >p$
by assumption, the r.h.s.'s of eqs.~(\ref{OL}) and~(\ref{OT}) are both
positive, so that $\Omega_L$ and $\Omega_T$  are real
functions of $\omega/p$.

Any solution of the mean field equations (\ref{const})--(\ref{OT}) depends
 parametrically on the wave-vector $p^\mu$, i.e. ${\cal A}^\mu_a=
{\cal A}^\mu_a(p\cdot x;p_\nu)$.
It can be easily verified that ${\cal A}^\mu_a$ is homogeneous
in $p^\mu$ of degree zero,
i.e.,  ${\cal A}^\mu_a(\lambda p\cdot x;\lambda p_\nu)
={\cal A}^\mu_a(p\cdot x;p_\nu)$ for arbitrary constant $\lambda$.
 Therefore, the solution depends only on
three parameters, for instance, $p^i/\omega$, with $i=1,\,2,\,3$.

At this stage, it is convenient to introduce three
 polarization vectors $\epsilon^\mu(p;s)$, with $s=1,\,2,\,3$
and $p_\mu\epsilon^\mu(p;s)=0$. We choose the vectors
$\epsilon^\mu(p;s=1,\,2)$  transverse   to $\hat {\bf p}$, i.e.,
\beq\label{tr}
\epsilon^\mu(p;s)\,=\,(0,{\bf e}_s({\bf p})),\qquad {\bf p}\cdot
{\bf e}_s({\bf p}) =0,\,\,\,\,{\bf e}_s({\bf p})\cdot {\bf e}_s^\prime({\bf p})
=\delta_{s s^\prime},\eeq
 while   \beq
\epsilon^\mu(p;3)\,=\,\frac{1}{\sqrt{\omega^2-p^2}}\,
(p,\omega{\bf\hat p}).\eeq
The  normalization is such that  $\epsilon (p;s)\cdot \epsilon (p;s^\prime)
\,=\,-\,\delta_{s s^\prime}$. The decomposition of ${\cal A}^\mu_a$
on the vectors $\epsilon^\mu(p;s)$ reads
\beq\label{phi}{\cal A}^\mu_a(z)\,=\,\sum_{s=1}^{3}
\epsilon^\mu(p;s)\,\phi^s_a(z),\eeq
where $\phi^s_a(z)$ are  new unknown functions.
In terms of these functions, the constraint (\ref{const}) reduces to
($\phi^s\equiv \phi^s_a\,T^a$) \beq\label{con1}
\sum_{s=1}^{3}\Bigl[\phi^s, \dot\phi^s\Bigr]\,=\,0.\eeq
This is satisfied, in particular, by field
configurations of the form
$\phi^s_a(z)\,=\,{\cal O}^s_a\,h_s(z)$ (no summation over $s$),
 with constant ${\cal O}^s_a$ and arbitrary  functions $h_s(z)$.
Indeed, for such fields, the color vectors $\{\phi^s_a(z)\}$
and $\{\dot\phi^s_a(z)\}$ ($s=1,\,2,\,3$) are  parallel in color space for
any $s$. In this letter we restrict ourselves to such configurations
for the color group   $SU(2)$ ($f^{abc}=\epsilon^{abc}$), and
 assume, for simplicity, that ${\cal O}^s_a=\delta^s_a$, i.e.,
\beq\label{AN}
{\cal A}^\mu(z)\,=\,\sum_{a=1}^{3}\epsilon^\mu(p;a)\,h_a(z)\,T^a.\eeq
The functions $h_a(z)$ ($a=1,\,2,\,3$)  satisfy
\beq\label{heq}
\Bigl\{(\omega^2 -p^2)\,\ddot h_a\,+\,g^2\,\Bigl(\sum_{b\not
 = a} h_b^2\Bigr)\,h_a\Bigr\}\,\epsilon^\mu(p;a)\,=\,j^\mu_a,\eeq
as follows from eqs.~(\ref{pweq1}). By  contracting
  these equations with $\epsilon_\mu(p;a)$ and using eqs.~(\ref{ind}),
we get
\beq\label{sys}
(\omega^2 -p^2)\,\ddot h_1\,+\,\Omega_T^2\,h_1\,+\,g^2\,(h_2^2+h_3^2)\,h_1
&=&0,\nonumber\\
(\omega^2 -p^2)\,\ddot h_2\,+\,\Omega_T^2\,h_2\,+\,g^2\,(h_1^2+h_3^2)\,h_2
&=&0,\nonumber\\
(\omega^2 -p^2)\,\ddot h_3\,+\,\Omega_L^2\,h_3\,+\,g^2\,(h_1^2+h_2^2)\,h_3
&=&0.\eeq
This system admits the following integral of the motion,
\beq\label{H}
{\cal H}\,=\,\frac{1}{2}\,(\omega^2-p^2)\sum_s \dot h_a^2+\frac{1}{2}\,
\Omega_T^2(h_1^2+h_2^2)+\frac{1}{2}\Omega_L^2h_3^2+
\frac{g^2}{2}\,\Bigl(h_1^2\,h_2^2+h_1^2\,h_3^2
+h_2^2\,h_3^2\Bigr),\eeq
which acts as an effective Hamiltonian for a point particle with coordinates
$h_a$ (the corresponding conjugate momenta being $\dot h_a$). The  potential
in (\ref{H}) is a positive,  strictly increasing function
 of the coordinates $h_a$. Accordingly,
the energy conservation prevents any trajectory $\{h_a(z)\}$
from getting too far away from the origin.

The limiting case $p=0$, $\omega\not = 0$, of the system (\ref{sys}),
corresponding to global color excitations of the plasma, has been investigated
in Refs. \cite{EMT,NAE}. In this limit, $\Omega_L=\Omega_T=\omega_p$,
and the system becomes symmetric with respect to permutations of the functions
$h_a$. It is then straigthforward to analytically construct particular periodic
solutions which take advantage of this symmetry. For example, the $p=0$ system
admits in-phase periodic oscillations of the type $h_1=h_2=h_3$.
More generally, numerical studies of the symmetric
system\cite{Matinyan81} showed that the  oscillations are
 quasi-periodic for small amplitude, but they become unstable as the
amplitude is increased. We expect these general properties to remain
valid for the asymmetric ($\Omega_T\not =\Omega_L$) system (\ref{sys}) as
well, and, in particular, the small amplitude abelian-like plasma waves
to be stable with respect to nonlinear effects. Note that,
 since the frequencies $\Omega_T$ and $\Omega_L$ are generally
incomensurable, we do not expect to find any periodic solution superposing
 {\it both} longitudinal and transverse waves.
However,  periodic solutions of the system (\ref{sys})  corresponding
 either to longitudinally, or to transversally polarized plane waves can
be constructed.

\setcounter{equation}{0}
\section {Longitudinal plane waves}

We set $h_1=h_2=0$  in eqs.~(\ref{sys}).
We get then a simple harmonic oscillator equation for $h_3$,
\beq\label{h3}
(\omega^2 -p^2)\,\ddot h_3\,+\,\Omega_L^2\,h_3\,=\,0,\eeq
which has the general solution $h_3(z)=C_1\,\cos \nu_L z +C_2\,\sin \nu_L z$,
with $\nu_L\equiv \Omega_L/\sqrt{\omega^2-p^2}$ and
 $C_1$, $C_2$ are  integration constants.
 It is convenient here to replace $p^\mu=(\omega,{\bf p})$
by $k^\mu \equiv \nu_L \,p^\mu= (k^0,{\bf k})$, and
 to set $h_L(k\cdot x)\equiv h_3(z)$.
Then, $\Omega_L^2= k^\mu k_\mu$ and $h_L$ satisfies
\beq\label{hL}\ddot h_L\,+\,h_L\,=\,0.\eeq
(The dots refer here to derivation with respect to $k\cdot x$.)
With these new notations, the longitudinal plane-waves that we consider
are of the form \beq\label{AL}
A_L^\mu(x)\,=\,\epsilon^\mu(k;3)\,h_L(k\cdot x)\,T^3,\eeq
where the frequency $k^0$ is related to the wave vector $k\equiv |{\bf k}|$
by
\beq\label{disl}
k^2\,=\,3\,\omega_p^2\,\biggl(Q\biggl(\frac{k^0}{k}\biggr)
\,-\,1\biggr).\eeq
This last equation, which is easily deduced from (\ref{OL}), is identical
to the dispersion equation of the abelian-like longitudinal
 modes\cite{Klimov81}. This is not surprising since the Ansatz
(\ref{AL}) led us effectively to a linear, abelian-like, problem.
In fact,  this property holds for any purely longitudinal plane-wave.
 Indeed, in that case,
 ${\cal A}^\mu_a(z) =
\epsilon^\mu(p;3)\,\phi^L_a(z)$, so that $[{\cal A}^\mu, {\cal A}^\nu]=0$
for any pair $(\mu,\nu)$ and the  general equations (\ref{pweq1})
 reduce to harmonic equations for each of the
functions $\phi_a^L(z)$. The  only manifestation of the non-abelian
structure lies in the constraint (\ref{const}), which becomes
 $[\phi^L, \dot \phi^L]=0$. A
 particular solution is $\phi_a^L(z)=q_a\,h_L(z)$, with constant $q_a$'s.
For $SU(2)$, this is the  only solution, and it is
equivalent to the Ansatz (\ref{AL}), up to a trivial global color rotation
(sending the color vector $q_a\,T^a$ onto the $T^3$ axis). For a larger
color group, other solutions will exist; e.g., for $SU(3)$ we can choose
$\phi^L(z)=\phi^L_3(z)\,T^3\,+\,\phi^L_8(z)\,T^8$, where $T^3$ and $T^8$
are the commuting generators of the group.

To complete our analysis of the longitudinal modes, we evaluate the
corresponding energy density and Poynting vector. We return to the particular
configuration (\ref{AL}) and denote by $k^0=\omega_l(k)$ the
solution  of the dispersion equation (\ref{disl})\cite{Klimov81}
 ($\omega_l(k)> k$ for any $k$, $\omega_l(0)=\omega_p$).
The associated field strengths are
 ${\bf E}_L(x)=-\hat {\bf k}\,\sqrt{\omega_l^2(k)-k^2}\,\dot h_L\,
T^3$  and ${\bf B}_L=0$.  By using these results, together with
eqs.~(\ref{enden})--(\ref{Poyn}), (\ref{ennon})--(\ref{Pnon}),
and the dispersion relation (\ref{disl}), we obtain
\beq\label{TL}
T^{00}_L(x)\,=\,\frac{1}{2}\Bigl(\omega_l^2(k)-k^2\Bigr)\,\biggl\{
\dot h_L^2\,+\,\biggl(\frac{3\omega_p^2}{\omega_l^2(k)-k^2}
\,-\,2\biggr)\, h_L^2\biggr\},\eeq
and \beq\label{SL}
{\bf S}_L(x)\,=\,\frac{3}{2}\,\frac{\omega_l(k)}{k}
\biggl\{\omega_p^2\,-\,\Bigl(\omega_l^2(k)-k^2\Bigr)\,\biggr\}h_L^2\,
\hat{\bf k}.\eeq As expected, these quantities satisfy the conservation law
\beq\label{consl}
\del_0\,T_L^{00}(x)\,+\,\del_i\,S_L^i(x)\,=\,0,\eeq
for $h_L$ satisfying (\ref{hL}). This last equation also clarifies the
physical interpretation of the conserved quantity ${\cal H}$, eq.~(\ref{H}).
Indeed, since the fields depend on $x^\mu$ only through $k\cdot x$, it follows
that $\del_0\,T_L^{00}\,=\,\omega_l(k)\,\dot T_L^{00}$ and
$\del_i\,S_L^i\,=\,-{\bf k}\cdot\dot{\bf S}_L\,=\,-\,(k/\omega_l(k))\del_0\,
\hat{\bf k}\cdot {\bf S}_L$. Thus, the conservation law (\ref{consl})
shows that the following quantity
\beq\label{TSL}
T_L^{00}(x)\,-\,\frac{k}{\omega_l(k)}\,\hat{\bf k}\cdot {\bf S}_L(x)\,=\,
\frac{1}{2}\Bigl(\omega_l^2(k)-k^2\Bigr)\,\Bigl(
\dot h_L^2\,+\, h_L^2\Bigr)\eeq
is constant along the trajectory. This is precisely the integral of the motion
 ${\cal H}_L$, as obtained from
eq.~(\ref{H}) in which we set $h_1=h_2=0$ and $\Omega_l^2=\omega_l^2(k)-k^2$.

\setcounter{equation}{0}
\section{Transverse plane waves}

We turn now to the more interesting case of purely transverse plane waves,
that is, we consider the system (\ref{sys}) for $h_3=0$. The  resulting two
equations,
\beq\label{tsys}
(\omega^2 -p^2)\,\ddot h_1\,+\,\Omega_T^2\,h_1\,+\,g^2\,h_2^2\,h_1
&=&0,\nonumber\\
(\omega^2 -p^2)\,\ddot h_2\,+\,\Omega_T^2\,h_2\,+\,g^2\,h_1^2\,h_2
&=&0,\eeq
are symmetric in $h_1$ and $h_2$. Again, it is convenient to
define $\nu_T\equiv \Omega_T/\sqrt{\omega^2-p^2}$ and
 $k^\mu\equiv \nu_T\,p^\mu$. Then, $\Omega_T^2=k^\mu k_\mu$ and from
(\ref{OT}) one easily obtains a dispersion equation for $k^0$,
\beq\label{dist}
k^2\,=\,\frac{3}{2}\,\omega_p^2\,\biggl[\frac{k_0^2}{k_0^2-k^2}\,-\,
Q\biggl(\frac{k^0}{k}\biggr)\biggr],\eeq
which is identical to that of the linear (or abelian)
 transverse modes\cite{Klimov81}. We denote by
$\omega_t(k)$ the positive solution ($\omega_t(k)> k$, $\omega_t(0)=\omega_p$).
The transverse plane waves that we are considering here are, in these
 notations,  of the form
\beq\label{AT}
A^0_T=0,\qquad {\bf A}_T(x)\,=\,h_T^1(k\cdot x)\,{\bf e}_1\,T^1\,+
\,h_T^2(k\cdot x)\,{\bf e}_2\,T^2,\eeq
where ${\bf e}_a\equiv {\bf e}_a({\bf k})$ (recall eq.~(\ref{tr})) and
$h_T^a(k\cdot x)\equiv h_a(z)$, for $a=1,\,2$.
The corresponding field strengths are
${\bf E}_T(x)\,=\,-\,\omega_t(k)\,\sum_{a=1,2}\, \dot h_T^a\,{\bf e}_a\,T^a$,
 and 
${\bf B}_T(x)\,=\,k\Bigl(\dot h_T^2\,{\bf e}_1\,T^2\,-\,
\dot h_T^1\,{\bf e}_2\,T^1\Bigr)\,+\,g\,h_T^1\,h_T^2\,\hat{\bf k}\,T^3$,
the overdots denoting here derivatives with respect to $k\cdot x$. Remark that
the three chromoelectric vectors ${\bf E}_a$ ($a=1,\,2,\,3$) are mutually
orthogonal, and the same is  true for the three chromomagnetic vectors
${\bf B}_a$ (of course, ${\bf E}_3 =0$). Furthermore, the electric and the
magnetic color vectors are orthogonal, ${\bf E}_a\cdot {\bf B}_a =0$ for any
color $a$. This is very much similar to the usual transverse electromagnetic
 waves.

The energy density and the Poynting vector corresponding to transverse
plane-waves are obtained by inserting the expressions for the field potentials
and strengths above
in eqs.~(\ref{enden})--(\ref{Poyn}), (\ref{ennon})--(\ref{Pnon}), and
 using the dispersion relation (\ref{dist}):
\beq\label{TT}
T^{00}_T(x)&=&\frac{1}{2}\Bigl(\omega_t^2(k)+k^2\Bigr)\,\Bigl[
\bigl(\dot h_T^1\bigr)^2\,+\,\bigl(\dot h_T^2\bigr)^2\Bigr]
\,+\,\frac{g^2}{2}\,\bigl(h_T^1\bigr)^2\,\bigl(h_T^2\bigr)^2\nonumber\\
&+&\frac{\omega_t^2(k)}{2}\,\biggl(\frac{3\omega_p^2}{\omega_t^2(k)-k^2}
\,-\,2\biggr)\Bigl[\bigl(h_T^1\bigr)^2\,+\,\bigl(h_T^2\bigr)^2\Bigr]\eeq
and \beq\label{ST}
{\bf S}_T(x)\,=\,{\bf k}\,\omega_t(k)\,\biggl\{
\bigl(\dot h_T^1\bigr)^2+\bigl(\dot h_T^2\bigr)^2+\frac{1}{2}\,
\biggl[1+3\,\frac{\omega_t^2(k)}{k^2}\biggl(\frac{
\omega_p^2}{\omega_t^2(k)-k^2}-1\biggr)\biggr]
\Bigl[\bigl(h_T^1\bigr)^2+\bigl(h_T^2\bigr)^2\Bigr]\biggr\}.
\nonumber\\ \eeq
One easily verifies that
\beq\label{HT}
T_T^{00}(x)\,-\,\frac{k}{\omega_t(k)}\,\hat{\bf k}\cdot {\bf S}_T(x)\,=\,
{\cal H}_T,\eeq
where the integral of the motion ${\cal H}_T$ is obtained by setting $h_3=0$ in
  eq.~(\ref{H}).

In order to look for solutions of the system (\ref{tsys}), we
define $h_T^a(k\cdot x)\equiv (\Omega_T/g)f_a(k\cdot x)$,
$a=1,\,2,$ and get the parameter-free system
\beq\label{fsys}
{\ddot f}_1(x)\,+\,\Bigl[1+\Bigl(f_2(x)\Bigr)^2\,\Bigr]\,f_1(x)&=&0,\nonumber
\\{\ddot f}_2(x)\,+\,\Bigl[1+\Bigl(f_1(x)\Bigr)^2\,\Bigr]\,f_2(x)&=&0,\eeq
  for the dimensionless functions $f_1$ and $f_2$.
The integral of the motion ${\cal H}_T$ reads
\beq\label{HTF}
{\cal H}_T&=&\frac{\Omega_T^4}{g^2}\,\frac{1}{2}\,\biggl(
{\dot f}_1^2+{\dot f}_2^2 + f_1^2+f_2^2 +f_1^2\,f_2^2\biggr)\equiv
\frac{\Omega_T^4}{g^2}\,\theta^2,\eeq
where $\Omega_T^2=\omega_t^2(k)-k^2$ and $\theta$ is a dimensionless parameter.
The Hamiltonian ${\cal H}_T$, to within the factor ${\Omega_T^4}/{g^2}$,
 is that of a system of two nonlinearly
 coupled harmonic oscillators with coordinates $f_1$ and $f_2$ and total energy
$\theta^2$. The system (\ref{fsys}) has been already analyzed in Refs.
\cite{NAE,EMT}, where analytic periodic solutions were constructed, as well as
in Refs. \cite{Matinyan81}, where the transition from regular
 to stochastic motion was investigated numerically.

The simplest solutions correspond to one-dimensional harmonic oscillations
along the space-color axes 1 or 2. For instance, if $f_2=0$, then
 $f_1$ satisfies ${\ddot f}_1\,+\,f_1=0$, with the
 general solution  $f_1\,=\,a_1\,\cos k\cdot x+a_2\,\sin k\cdot x$, and
 $a_1^2 + a_2^2= 2\theta^2$. This is time periodic,
with period ${\cal T}_0=2\pi/\omega_t(k)$.

Non-linear, but still one-dimensional, periodic solutions are analytically
obtained with the Ansatz  $f_1=\pm f_2\equiv f$.
These  solutions describe in or out of phase
oscillations along the  space-color directions 1 and 2.
The function $f(k\cdot x)$ satisfies the nonlinear equation
\beq\label{fnon}
{\ddot f}\,+\,f\,+\,f^3&=&0.\eeq
A particular solution to (\ref{fnon}) is
\beq\label{fsol}
f(x)\,=\,f_\theta\,{\rm cn}\Bigl((2\theta^2+1)^{1/4}\,k\cdot x;
\kappa\Bigr),\eeq
 corresponding to the initial conditions
$f(0)=f_\theta$ and ${\dot f}(0)=0$. Here,
 ${\rm cn}(u;\kappa)$ is the Jacobi elliptic cosine of argument $u$ and
modulus $\kappa$, with
\beq\label{k}
\kappa\equiv \frac {1}{\sqrt 2}\,\biggl(1\,-\,\frac{1}{\sqrt {2\theta^2+1}}
\biggr)^{1/2},\eeq
 and the amplitude $f_\theta$ is related to the conserved quantity $\theta$  by
$f_\theta\,=\,\Bigl({\sqrt {2\theta^2+1}}\,-\,1\Bigr)^{1/2}$.
The solution (\ref{fsol}) is time periodic, with period
\beq\label{T}
{\cal T}\,=\,\frac{4}{\omega_t(k)}\,\frac{1}{(2\theta^2+1)^{1/4}}\,K(\kappa),
\eeq where $K(\kappa)$ is the complete elliptic integral of modulus $\kappa$.
Since $|{\bf A}_T| \simle T$, $\theta\simle 1$, and
  ${\cal T}_\theta$ remains of order of  ${\cal T}_0\equiv 2\pi/\omega_t(k)$.

\setcounter{equation}{0}
\section{Conclusions}

We have studied here the generalized Yang-Mills equations in the hot
 quark-gluon plasma for particular plane-wave fields. For the color
group $SU(2)$ and for an appropriate
Ansatz  showing correlations between coordinate and color spaces,
the problem reduces to an effective  mechanical system, with
three degrees of freedom and nonlinear couplings. Generally, this system
 describes  superpositions of longitudinal and transverse plane waves,
with components along all the three color directions. Simple periodic
solutions were  obtained analytically for the particular case where
the longitudinal and the transverse plane waves decouple. In contrast
to what happens at zero temperature, here the plane-wave solutions cannot
be obtained by simply performing  boost transformations on the corresponding
solutions in the comoving frame\cite{Baseyan79}.
 This is so because the mean field
equations and their solutions are written in the rest frame of the plasma.

\end{document}